\documentclass[12pt]{article}
\usepackage{amsfonts}
\usepackage{amsmath}
\usepackage{extarrows}

\usepackage{mathrsfs}
\title{\textbf{Quasi-periodic solutions of the Heisenberg hierarchy}}
\date{}
\author{\\Xianguo Geng, Zhu Li\thanks{Corresponding author. E-mail address: lizhu1813@163.com}, Liang Guan\\
{\small{\sl School of Mathematics and Statistics, Zhengzhou University, }}\\
{\small{\sl 100 Kexue Road, Zhengzhou, Henan 450001, People's
Republic of China}}}
\begin{document} \maketitle

\begin{abstract}
The Heisenberg hierarchy and its Hamiltonian structure are derived
respectively by virtue of the zero curvature equation and the trace
identity. With the help of the Lax matrix we introduce an algebraic
curve $\mathcal{K}_{n}$ of arithmetic genus $n$, from which we
define meromorphic function $\phi$ and straighten out all of the
flows associated with the Heisenberg hierarchy under the Abel-Jacobi
coordinates. Finally, we achieve the explicit theta function
representations of solutions for the whole Heisenberg hierarchy as a
result of the asymptotic properties of $\phi$.


\end{abstract}

\section{Introduction}
It is well known that seeking quasi-periodic solutions of soliton
equations is a very important topic in soliton theory because
soliton equations describe various nonlinear phenomena in natural
and applied sciences such as fluid dynamics, plasma physics, solid
state physics, optical fibers, acoustics, mechanics, biology and
mathematical finance. With the development of soliton theory, many
approaches were developed from which quasi-periodic solutions for
several soliton equations have been obtained in Refs. [1-21], such as
the KdV, mKdV, nonlinear Schr\"odinger, sine-Gordon, Toda lattice
and Camassa-Holm equations, etc.

In this paper, we would like to construct quasi-periodic solutions
of the Heisenberg  hierarchy by means of the methods in Refs.
[15,21]. Ferromagnetic chain equation was first proposed in 1935 by
Landau and Lifshitz when studying the dispersive theory for magnetic
conductivity in magnetic materials [22]. It is an important dynamical
equation, which has coherent and chaotic structures depending on the
nature of magnetic interactions, and frequently appears in
condensation physics, quantum physics and other physics
fields [23-25]. The Lax integration of the continuous Heisenberg spin
chain equation are studied by Takhtajan through the inverse
scattering transform method in 1977 [26]. Almost at the same time the
single-soliton solution of Heisenberg spin chain in the isotropic
case are obtained by Tjon and Wright [27]. The classical solutions of
the continuous Heisenberg spin chain has been obtained by Jevicki
and Papanicolaou using a path integral formalism that allows for
semi-classical quantization of systems with spin degrees of freedom
in 1979 [28]. Soon after, an explicit expression is obtained for the
Miura transformation which maps the solutions of the continuous
Anisotropic Heisenberg Spin Chain on solutions of the Nonlinear
Schr\"odinger equation  by Quispel and Capel in 1983 [29]. Li and
Chen gave the higher order Heisenberg spin chain equations, and they
proved that these evolution equations are equivalent to the
evolution equation of AKNS type in 1986 [30]. Afterwards, The role of
nonlocal conservation laws and the corresponding charges are
analyzed in the supersymmetric Heisenberg spin chain in 1994 [31].
Cao discussed the parametric representation of the finite-band
solution of the Heisenberg equation [32]. The algebraic Bethe ansatz
equation has been set up for an open Heisenberg spin chain having an
impurity of a different type of spin [33]. Qiao gave the involutive
solutions of the higher order Heisenberg spin chain equations by
virtue of the spectral problem nonlinearization method [34].
Recently, Du derived the Poisson reduction and Lie-Poisson structure
for the nonlinearized spectral problem of the Heisenberg hierarchy
by the method of invariants [35]. Wang and Zanardi showed that in
one-dimensional isotropic Heisenberg model two-qubit thermal
entanglement and maximal violation of Bell inequalities are directly
related with a thermodynamical state function [36]. Wang studied the
Darboux transformation for the Heisenberg hierarchy and constructed
explicit soliton solutions for the hierarchy by using the Darboux
transformation [37]. Guo and others proved the existence of periodic
weak solutions to the classical one-dimensional isotropic
biquadratic Heisenberg spin chain in 2007 [38]. Its and Korepin
considered the XY quantum spin chain in a transverse magnetic field
in 2010 [39]. Li and others investigated the gauge transformation
between the first-order nonisospectral and isospectral Heisenberg
hierarchies [40]. Miszczak and others studied a quantum version of a
penny flip game in the Heisenberg model [41].

The paper is structured as follows. In section 2, with the aid of
the zero-curvature equation and the trace identity we derive the
Heisenberg hierarchy and its Hamiltonian structure. In section 3,
the nonlinear recursion relations of the homogeneous coefficients
are given based on a Lax matrix and an algebraic curve
$\mathcal{K}_{n}$ of arithmetic genus $n$. In section 4, we first
get the Dubrovin-type equations of the elliptic variables, then we
straightened out all the flows of the Heisenberg hierarchy under the
Abel-Jacobi coordinates. In section 5, we constructed the
quasi-periodic solutions of the whole Heisenberg hierarchy by use of
the Riemann theta functions according to the asymptotic properties
 of the meromorphic function $\phi$.

\section{The Heisenberg hierarchy and its Hamiltonian structure}\setcounter{equation}{0}
\hspace{0.5cm} In this section, we shall derive the Heisenberg
hierarchy associated with the $2\times2$ spectral problem [26]
\begin{equation}
      \begin{split}
                   \varphi_{x}=U\varphi,\quad
                                \varphi=\left(
                                             \begin{array}{c}
                                                 \varphi_{1}\\
                                                 \varphi_{2}
                                             \end{array}
                                         \right),\quad
                  U=\lambda \left(
                                 \begin{array}{cc}
                                          w    &   u    \\
                                          v    &   -w   \\
                                 \end{array}
                            \right),\\
      \end{split}
\end{equation}
where $w^{2}+uv=1$, $u$ and $v$ are two potentials, $\lambda$ is a
constant spectral parameter.  To this end, we solve the stationary
zero-curvature equation
\begin{equation}
     V_{x}-[U,V]=0,\quad V=\lambda\left(
                                         \begin{array}{cc}
                                               -\frac{1}{2}w A       &   B-\frac{1}{2}uA     \\
                                               C-\frac{1}{2}vA       &   \frac{1}{2}w A      \\
                                         \end{array}
                                  \right),
\end{equation}
which is equivalent to
\begin{equation}
         \begin{split}
                     \frac{1}{2}(w A)_{x}+\lambda(uC-vB)=0,    \\
                     B_{x}-\frac{1}{2}(u A)_{x}-2\lambda w B=0,\\
                     C_{x}-\frac{1}{2}(v A)_{x}+2\lambda w C=0,\\
         \end{split}
\end{equation}
where
\begin{equation}
                    A=\sum\limits_{j\geq 0}a_{j-1}\lambda^{-j},\quad
                    B=\sum\limits_{j\geq 0}b_{j-1}\lambda^{-j},\quad
                    C=\sum\limits_{j\geq 0}c_{j-1}\lambda^{-j}.
\end{equation}
A direct calculation shows that  (2.3) and (2.4) imply the Lenard
recursion equations
\begin{equation}
        KL_{j-1}=JL_{j},\quad JL_{-1}=0,\quad L_{j}=(c_{j},b_{j},a_{j})^{T}
\end{equation}
in which $K$ and $J$ are two operators defined by
\begin{equation}
         K=\begin{pmatrix}
                  0               & \partial      & -\frac{1}{2} \partial u    \\
                  \partial        & 0             & -\frac{1}{2} \partial v    \\
                  u\partial       & v\partial     & -\partial                  \\
            \end{pmatrix},\quad
         J=\begin{pmatrix}
                  0               & 2w            & 0           \\
                  -2w             & 0             &0            \\
                  u\partial       & v\partial     & -\partial   \\
            \end{pmatrix}.
\end{equation}
We here take
\begin{equation}
          L_{-1}=(0,0,-2)^{T}
\end{equation}
as a starting point. It is easy to see that $\mbox{Ker}J=\{\bar{c}_0
L_{-1}\ |\ \bar{c}_0\in{\mathbb R}\}$. Then $L_{j}$ is uniquely
determined by the recursion relation (2.5) up to a term
const.$L_{-1}$, which is always assumed to be zero. The first two
members are
\begin{equation}
          L_{0}=\frac{1}{2w}\left(\begin{array}{cc}
                                              -v_{x}         \\
                                              u_{x}          \\
                                              u_{x}v-uv_{x}
                                        \end{array}
                                 \right),\quad
          L_{1}=\frac{1}{4w}\left(\begin{array}{cc}
                                              v_{xx}w-vw_{xx}       \\
                                              u_{xx}w-uw_{xx}       \\
                                              -2w_{xx}-3w(u_{x}v_{x}+w_{x}^{2})
                                       \end{array}
                                 \right).
\end{equation}

Assume that the time evolution of the eigenfunction $\varphi$ obeys
the differential equation
\begin{equation}
            \varphi_{t_{m}}=V^{(m)}\varphi,\quad V^{(m)}=\left(\begin{array}{cc}
                                                                           V_{11}^{(m)} & V_{12}^{(m)}\\
                                                                           V_{21}^{(m)} & -V_{11}^{(m)}
                                                         \end{array}\right),
\end{equation}
where $V_{11}^{(m)},V_{12}^{(m)},V_{21}^{(m)}$ are polynomials of
the spectral parameter $\lambda$ with
\begin{equation}
             \begin{split}
                          &V_{11}^{(m)}=\sum\limits_{j=0}^{m}\left(-\frac{1}{2}w a_{j-1}\right)\lambda^{m+1-j},\\
                          &V_{12}^{(m)}=\sum\limits_{j=0}^{m}\left(b_{j-1}-\frac{1}{2}u a_{j-1}\right)\lambda^{m+1-j},\\
                          &V_{21}^{(m)}=\sum\limits_{j=0}^{m}\left(c_{j-1}-\frac{1}{2}v a_{j-1}\right)\lambda^{m+1-j}.
             \end{split}
\end{equation}
Then the compatibility condition of (2.1) and (2.9) yields the zero
curvature equation, $U_{t_{m}}-V^{(m)}_{x}+[U,V^{(m)}]=0,$ which is
equivalent to the hierarchy of nonlinear evolution equations
\begin{equation}
               u_{t_{m}}=2w b_{m},\quad
               v_{t_{m}}=-2w c_{m}.
\end{equation}
The first two nontrival members in the hierarchy are as follows
\begin{equation}
               u_{t_{1}}=\frac{1}{2}(u_{xx}w-uw_{xx}),\quad
               v_{t_{1}}=\frac{1}{2}(w_{xx}v-w v_{xx}),\\
\end{equation}
\begin{equation}
               u_{t_{2}}=\frac{1}{4}u_{xxx}+\frac{3}{8}(uu_{x}v_{x}+uw_{x}^{2})_{x},\quad
               v_{t_{2}}=\frac{1}{4}v_{xxx}+\frac{3}{8}(vu_{x}v_{x}+vw_{x}^{2})_{x}.
\end{equation}

In the following we derive the Hamiltonian structure of the
hierarchy (2.11). A direct calculation gives
\begin{equation}
        \mathrm{tr}\left(V\frac{\partial U}{\partial\lambda}\right)=\lambda(uC+vB-A),\quad
        \mathrm{tr}\left(V\frac{\partial U}{\partial u}\right)=\lambda^{2}C,\quad
        \mathrm{tr}\left(V\frac{\partial U}{\partial v}\right)=\lambda^{2}B.
\end{equation}
Substituting (2.14) into the trace identity [42] leads to
\begin{equation}
    \left(
          \begin{array}{c}
                \frac{\delta}{\delta u} \\
                \frac{\delta}{\delta v}\\
          \end{array}
    \right)(\lambda(uC+vB-A))
    =\lambda^{-\gamma}\left(\frac{\partial}{\partial\lambda}\lambda^{\gamma}\left(
                                                                                  \begin{array}{c}
                                                                                            \lambda^{2}C \\
                                                                                            \lambda^{2}B \\
                                                                                  \end{array}
                                                                             \right)
                      \right).
\end{equation}
Comparing the coefficient of the $\lambda^{-n}$ in (2.15) yields
\begin{equation}
           \left(
                 \begin{array}{c}
                          \frac{\delta}{\delta u} \\
                          \frac{\delta}{\delta v}\\
                 \end{array}
           \right)
           \left(uc_{n}+vb_{n}-a_{n}\right)=(\gamma-n+1)\left(
                                                             \begin{array}{c}
                                                                         c_{n} \\
                                                                         b_{n}\\
                                                             \end{array}
                                                        \right).
\end{equation}
Let $n=1$ in (2.16) and find that $\gamma=-1$, so we have
\begin{equation}
         \frac{\delta H_{n}}{\delta \eta}=\left(
                                             \begin{array}{c}
                                                       c_{n} \\
                                                       b_{n} \\
                                             \end{array}
                                       \right),\quad
         \eta=\left(
                   \begin{array}{c}
                            u \\
                            v \\
                   \end{array}
             \right),\quad
         H_{n}=\frac{a_{n}-uc_{n}-vb_{n}}{n}.
\end{equation}
Therefore, the Hamiltonian structure of the Heisenberg hierarchy
(2.11) is as follows
\begin{equation}
    \left(\begin{array}{c}
                 u \\
                 v \\
         \end{array}
    \right)_{t_{n}}=\bar{J}\frac{\delta H_{n}}{\delta \eta},\quad \bar{J}=\left(
                                                                             \begin{array}{cc}
                                                                                       0        &  2w    \\
                                                                                       -2w      &  0     \\
                                                                             \end{array}
                                                                       \right).
\end{equation}

\section{Nonlinear recursion relations}
\setcounter{equation}{0} \hspace{0.5cm}
 Let $\chi=(\chi_{1},\chi_{2})^{T}$ and $\psi=(\psi_{1},\psi_{2})^{T}$ be
two basic solutions of (2.1) and (2.9). We introduce a Lax matrix
\begin{equation}
          W=\frac{1}{2}(\chi\psi^{T}+\psi\chi^{T})\left(
                                                        \begin{array}{cc}
                                                                   0 & -1 \\
                                                                   1 & 0  \\
                                                        \end{array}
                                                  \right)
          =\lambda\left(
                       \begin{array}{cc}
                                   G & F  \\
                                   H & -G \\
                       \end{array}
                  \right)
\end{equation}
which satisfies the Lax equations
\begin{equation}
      W_{x}=[U,W],\quad W_{t_{m}}=[V^{(m)},W].
\end{equation}
Therefore, det$W$ is a constant independent of $x$ and $t_{m}$.
Equation (3.2) can be written as
\begin{equation}
       \begin{split}
               G_{x}&=\lambda (uH-vF),\\
               F_{x}&=2\lambda (w F-uG),\\
               H_{x}&=2\lambda (vG-w H),
       \end{split}
\end{equation}
and
\begin{equation}
        \begin{split}
               G_{t_{m}}&=V_{12}^{(m)}H-V_{21}^{(m)}F,\\
               F_{t_{m}}&=2(V_{11}^{(m)}F-V_{12}^{(m)}G),\\
               H_{t_{m}}&=2(V_{21}^{(m)}G-V_{11}^{(m)}H).
       \end{split}
\end{equation}
Suppose functions $F$, $G$ and $H$ are finite-order polynomials in
$\lambda$
\begin{equation}
        \begin{split}
              G=\sum\limits_{j=0}^{n+1}G_{j-1}\lambda^{n+1-j},\quad
              F=\sum\limits_{j=0}^{n+1}F_{j-1}\lambda^{n+1-j},\quad
              H=\sum\limits_{j=0}^{n+1}H_{j-1}\lambda^{n+1-j},
        \end{split}
\end{equation}
where
\begin{equation}
         G_{j-1}=-\frac{1}{2}w g_{j-1},\quad
         F_{j-1}=f_{j-1}-\frac{1}{2}ug_{j-1},\quad
         H_{j-1}=h_{j-1}-\frac{1}{2}vg_{j-1}.
\end{equation}
 Substituting (3.5) and (3.6) into (3.3) yields
\begin{align}
         &KE_{j-1}=JE_{j},\quad JE_{-1}=0,\\
         &KE_{n}=0,
\end{align}
where $E_{j}=(h_{j},f_{j},g_{j})^{T}$, $-1\leq j\leq n$. It is easy
to see that the equation $JE_{-1}=0$ has the general solution
\begin{equation}
         E_{-1}=\alpha_{-1}(0,0,-2)^{T}.
\end{equation}
Without loss of generality, let $\alpha_{-1}=1$. If we take (3.9) as
a starting point, then $E_{j}$ can be recursively determined by the
relation (3.7). Acting with the operator $(J^{-1}K)^{k+1}$ upon
(3.9), we obtain from (3.7) and (2.5) that
\begin{equation}
         E_{k}=\sum\limits_{j=0}^{k+1}\alpha_{j-1}L_{k-j},\quad -1\leq k\leq n,
\end{equation}
where $\alpha_{0},\alpha_{1},\ldots,\alpha_{k}$ are constants of
integration. The first two members in (3.10) are
\begin{equation}
   E_{0}=\left(
              \begin{split}
                      &\quad\quad  -\frac{v_{x}}{2w}\\
                      &\quad\quad\quad  \frac{u_{x}}{2w}\\
                      &\frac{u_{x}v-uv_{x}}{2w}-2\alpha_{0}\\
              \end{split}
         \right),
\end{equation}
\begin{equation}
   E_{1}=\left(
             \begin{split}
                     &\quad\quad\quad\quad   \frac{v_{xx}w-vw_{xx}}{4w}-\alpha_{0}\frac{v_{x}}{2w}\\
                     &\quad\quad\quad\quad   \frac{u_{xx}w-uw_{xx}}{4w}+\alpha_{0}\frac{u_{x}}{2w}\\
                     &\frac{-2w_{xx}-3w(u_{x}v_{x}+w_{x}^{2})}{4w}+\alpha_{0}\frac{u_{x}v-uv_{x}}{2w}-2\alpha_{1}
             \end{split}
         \right).
\end{equation}
Since det$W$ is a $(2n+4)$th-order polynomial in $\lambda$, whose
coefficients are constants independent of $x$ and $t_{m}$, we have
\begin{equation}
          -\mathrm{det}W=\lambda^{2}(G^{2}+FH)=\lambda^{2}\prod\limits_{j=1}^{2n+2}(\lambda-\lambda_{j})=\lambda^{2}R(\lambda),
\end{equation}
one is naturally led to introduce the hyperelliptic curve
$\mathcal{K}_{n}$ of arithmetic genus $n$ defined by
\begin{equation}
          \mathcal{K}_{n}:y^{2}-R(\lambda)=0.
\end{equation}
The curve $\mathcal{K}_{n}$ can be compactified by joining two
points at infinity, $P_{\infty\pm}$, where $P_{\infty+}\neq
P_{\infty-}$. For notational simplicity the compactification of the
curve $\mathcal{K}_{n}$ is also denoted by $\mathcal{K}_{n}$. Here
we assume that the zeros $\lambda_{j}$ of $R(\lambda)$ in (3.13) are
mutually distinct, which means $\lambda_{j}\neq\lambda_{k}$, for
$j\neq k$, $1\leq j,k\leq 2n+2$. Then the hyperelliptic curve
$\mathcal{K}_{n}$ becomes nonsingular.

From the following lemma, we can explicitly represent
$\alpha_l(-1\leq l\leq n)$ by the constants
$\lambda_1,\ldots,\lambda_{2n+2}$.

\textbf{Lemma 3.1.}
\begin{equation}
         \alpha_l=c_l(\underline{\Lambda}),\qquad l=-1,\ldots,n,
\end{equation}
where
\begin{equation}
         \begin{split}
                \underline{\Lambda}&=(\lambda_1,\ldots,\lambda_{2n+2}),\quad
                c_{-1}(\underline{\Lambda})=1,\quad
                c_0(\underline{\Lambda})=-\frac{1}{2}\sum\limits_{j=1}^{2n+2}\lambda_j,\ldots,\\
                c_l(\underline{\Lambda})&=\sum\limits^{l+1}_{\stackrel{j_1,\ldots,j_{2n+2}=0}{j_1+\ldots+j_{2n+2}=l+1}}\frac{(2j_1)!\ldots(2j_{2n+2})!\lambda_1^{j_1}\ldots\lambda_{2n+2}^{j_{2n+2}}}{2^{2l+2}(j_1!)^2\ldots(j_{2n+2}!)^2(2j_1-1)\ldots(2j_{2n+2}-1)}.
         \end{split}
\end{equation}

\textbf{Proof.} Let
\begin{equation}
       \hat{F}_j=F_j|_{\alpha_0=\ldots=\alpha_j=0},\quad
       \hat{H}_j=H_j|_{\alpha_0=\ldots=\alpha_j=0},\quad
       \hat{G}_j=G_j|_{\alpha_0=\ldots=\alpha_j=0}.
\end{equation}
It will be convenient to introduce the notion of a degree,
$\deg(.)$, to effectively distinguish between homogeneous and
nonhomogeneous quantities. Define
\begin{equation}
       \deg(u)=0,\quad
       \deg(v)=0,\quad
       \deg(w)=0,\quad
       \deg(\partial_x)=1,
\end{equation}
thus from (3.7) and (3.10) it can be implied that
\begin{equation}
      \deg(\hat{F}_k)=k+1,\quad
      \deg(\hat{H}_k)=k+1,\quad
      \deg(\hat{G}_k)=k+1,\quad k\in\mathbb{N}_0\cup\{-1\}.
\end{equation}
Temporarily fixed the branch of $R(\lambda)^{1/2}$ as
$\lambda^{n+1}$ near infinity, $R(\lambda)^{-1/2}$ has the following
expansion
\begin{equation}
      R(\lambda)^{-1/2}\mathop{=}\limits_{\lambda\rightarrow\infty}\sum_{l=0}^{\infty}\hat{c}_{l-1}(\underline{\Lambda})\lambda^{-n-1-l},
\end{equation}
where
\begin{equation}
         \begin{split}
                    &\underline{\Lambda}=(\lambda_1,\ldots,\lambda_{2n+2}),\quad
                    \hat{c}_{-1}(\underline{\Lambda})=1,\quad
                    \hat{c}_0(\underline{\Lambda})=\frac{1}{2}\sum\limits_{j=1}^{2n+2}\lambda_j,\ldots,\\
                    &\hat{c}_l(\underline{\Lambda})=\sum\limits^{l+1}_{\stackrel{j_1,\ldots,j_{2n+2}=0}{j_1+\ldots+j_{2n+2}=l+1}}\frac{(2j_1)!\ldots(2j_{2n+2})!\lambda_1^{j_1}\ldots\lambda_{2n+2}^{j_{2n+2}}}{2^{2l+2}(j_1!)^2\ldots(j_{2n+2}!)^2}.
         \end{split}
\end{equation}
Dividing $F(\lambda)$, $H(\lambda)$, $G(\lambda)$ by
$R(\lambda)^{1/2}$ near infinity respectively, we obtain
\begin{equation}
             \begin{split}
                        &\frac{F(\lambda)}{R(\lambda)^{1/2}}\mathop{=}\limits_{\lambda\rightarrow\infty}\left(\sum_{l=0}^{n+1}F_{l-1}\lambda^{n+1-l}\right)\left(\sum_{l=0}^{\infty}\hat{c}_{l-1}(\underline{\Lambda})\lambda^{-n-1-l}\right)=\sum_{l=0}^{\infty}\check{F}_{l-1}\lambda^{-l},\\
                        &\frac{H(\lambda)}{R(\lambda)^{1/2}}\mathop{=}\limits_{\lambda\rightarrow\infty}\left(\sum_{l=0}^{n+1}H_{l-1}\lambda^{n+1-l}\right)\left(\sum_{l=0}^{\infty}\hat{c}_{l-1}(\underline{\Lambda})\lambda^{-n-1-l}\right)=\sum_{l=0}^{\infty}\check{H}_{l-1}\lambda^{-l},\\
                        &\frac{G(\lambda)}{R(\lambda)^{1/2}}\mathop{=}\limits_{\lambda\rightarrow\infty}\left(\sum_{l=0}^{n+1}G_{l-1}\lambda^{n+1-l}\right)\left(\sum_{l=0}^{\infty}\hat{c}_{l-1}(\underline{\Lambda})\lambda^{-n-1-l}\right)=\sum_{l=0}^{\infty}\check{G}_{l-1}\lambda^{-l},
             \end{split}
\end{equation}
for some coefficients $\check{F}_{l-1}$, $\check{H}_{l-1}$,
$\check{G}_{l-1}$ to be determined next. Noticing (3.3) and (3.13),
we get
\begin{equation}
         \begin{split}
                 \displaystyle-\frac{F_{xx}F}{2\lambda^{2}u^2}+\frac{F_x^2}{4\lambda^{2}u^2}+\frac{u_x}{2\lambda^{2}u^3}FF_x+F^2\left(\frac{1}{u^2}+\frac{w_xu-w u_x}{\lambda u^3}\right)=R(\lambda),\\
                 \displaystyle-\frac{H_{xx}H}{2\lambda^{2}v^2}+\frac{H_x^2}{4\lambda^{2}v^2}+\frac{v_x}{2\lambda^{2}v^3}HH_x+H^2\left(\frac{1}{v^2}+\frac{w v_x-w_xv}{\lambda
                 v^3}\right)=R(\lambda),
         \end{split}
\end{equation}
and
\begin{equation}
         \begin{split}
                 \displaystyle-\frac{1}{2\lambda u}GF_x+\frac{w}{u}GF+FH=R(\lambda),\\
                 \displaystyle\frac{1}{2\lambda v}GH_x+\frac{w}{v}GH+FH=R(\lambda),\\
                 \displaystyle\frac{1}{\lambda u}FG_x+\frac{v}{u}F^2+G^2=R(\lambda).
         \end{split}
\end{equation}
Respectively substituting (3.22) into (3.23), (3.13) and (3.24),
comparing the coefficients of $\lambda$ with the same power, we
arrive at the following recursive relations
\begin{equation}
          \begin{split}
                 \check{F}_{k-1}=&-\displaystyle\frac{1}{2u}\displaystyle\left\{\sum_{l=0}^{k-2}\left(-\frac{1}{2}\check{F}_{l-1,xx}\check{F}_{k-3-l}+\frac{1}{4}\check{F}_{l-1,x}\check{F}_{k-3-l,x}+\frac{u_x}{2u}\check{F}_{l-1,x}\check{F}_{k-3-l}\right)\right.\\
                                 &\left.+\displaystyle\sum_{l=1}^{k-1}\check{F}_{l-1}\check{F}_{k-1-l}+\sum_{l=0}^{k-1}\frac{w_xu-w u_x}{u} \check{F}_{l-1}\check{F}_{k-2-l}\right\},
          \end{split}
\end{equation}
\begin{equation}
          \begin{split}
                 \check{H}_{k-1}=&-\displaystyle\frac{1}{2v}\displaystyle\left\{\sum_{l=0}^{k-2}\left(-\frac{1}{2}\check{H}_{l-1,xx}\check{H}_{k-3-l}+\frac{1}{4}\check{H}_{l-1,x}\check{H}_{k-3-l,x}+\frac{v_x}{2v}\check{H}_{l-1,x}\check{H}_{k-3-l}\right)\right.\\
                                 &\left.+\displaystyle\sum_{l=1}^{k-1}\check{H}_{l-1}\check{H}_{k-1-l}+\sum_{l=0}^{k-1}\frac{w v_x-w _xv}{v} \check{H}_{l-1}\check{H}_{k-2-l}\right\},
          \end{split}
\end{equation}
\begin{equation}
          \check{G}_{k-1}=-\displaystyle\frac{1}{2w}\sum_{l=0}^{k}\check{F}_{l-1}\check{H}_{k-1-l}-\frac{1}{2w}\sum_{l=1}^{k-1}\check{G}_{l-1}\check{G}_{k-1-l}
\end{equation}
for $k\geq2$, and relations
\begin{equation}
          \check{F}_{k-1,x}=-\displaystyle\frac{1}{w}\sum_{l=0}^{k-1}\check{F}_{l-1,x}\check{G}_{k-1-l}+\sum_{l=0}^{k+1}\left(2\check{F}_{l-1}\check{G}_{k-l}+\frac{2u}{w}\check{F}_{l-1}\check{H}_{k-l}\right),\\
\end{equation}
\begin{equation}
          \check{H}_{k-1,x}=-\displaystyle\frac{1}{w}\sum_{l=0}^{k-1}\check{H}_{l-1,x}\check{G}_{k-1-l}-\sum_{l=0}^{k+1}\left(2\check{H}_{l-1}\check{G}_{k-l}+\frac{2v}{w}\check{F}_{l-1}\check{H}_{k-l}\right),
\end{equation}
\begin{equation}
          \check{G}_{k-1,x}=-\displaystyle\frac{1}{u}\sum_{l=0}^{k-1}\check{G}_{l-1,x}\check{F}_{k-1-l}-\sum_{l=0}^{k+1}\left(\check{G}_{l-1}\check{G}_{k-l}+\frac{v}{u}\check{F}_{l-1}\check{F}_{k-l}\right)
\end{equation}
for $k\geq1$ and
\begin{equation}
          \begin{split}
                  &\check{F}_{-1}=u,\quad         \check{F}_0=\frac{1}{2}(w u_x-w_xu),\\
                  &\check{H}_{-1}=v,\quad         \check{H}_0=\frac{1}{2}(w_x v-w v_x),\\
                  &\check{G}_{-1}=w,\quad         \check{G}_0=\frac{1}{4}(uv_x-u_xv).
          \end{split}
\end{equation}
The signs of $\check{F}_{-1}$, $\check{H}_{-1}$ and $\check{G}_{-1}$
have been chosen such that $\check{F}_{-1}=\hat{F}_{-1}$,
$\check{H}_{-1}=\hat{H}_{-1}$ and $\check{G}_{-1}=\hat{G}_{-1}$.
Moreover, we can prove inductively using the nonlinear recursion
relations (3.25)-(3.27) and (3.31) that
\begin{equation}
        \deg(\check{F}_k)=k+1,\quad
        \deg(\check{H}_k)=k+1,\quad
        \deg(\check{G}_k)=k+1,\quad
        k\in\mathbb{N}_0\cup\{-1\}.
\end{equation}
It can be proved inductively that
\begin{equation}
        \begin{split}
                 &\check{F}_{k-1,x}+2u\check{G}_{k}=2w\check{F}_{k},\\
                 &\check{H}_{k-1,x}-2v\check{G}_{k}=-2w\check{H}_{k},\\
                 &\check{G}_{k-1,x}=u\check{H}_{k}-v\check{F}_{k},
        \end{split}
\end{equation}
for $k\in\mathbb{N}_0\cup\{-1\}$. In fact, suppose that for
arbitrary $l$, $-1\leq l\leq k-2$, we have
\begin{equation}
        \begin{split}
                 &\check{F}_{l,x}+2u\check{G}_{l+1}=2w\check{F}_{l+1},\\
                 &\check{H}_{l,x}-2v\check{G}_{l+1}=-2w\check{H}_{l+1},\\
                 &\check{G}_{l,x}=u\check{H}_{l+1}-v\check{F}_{l+1},
        \end{split}
\end{equation}
then with the help of (3.27), (3.28) and (3.31), it can be
calculated out that
\begin{equation}
        \begin{split}
                  &\check{F}_{k-1,x}+2u\check{G}_{k}\\
                  &=-\displaystyle\frac{1}{w}\sum_{l=0}^{k-1}\check{F}_{l-1,x}\check{G}_{k-1-l}+\sum_{l=0}^{k+1}\left(2\check{F}_{l-1}\check{G}_{k-l}+\frac{2u}{w}\check{F}_{l-1}\check{H}_{k-l}\right)+2u\check{G}_{k}\\
                  &=-\displaystyle\frac{1}{w}\sum_{l=0}^{k-1}(2w\check{F}_{l}-2u\check{G}_{l})\check{G}_{k-1-l}+\sum_{l=0}^{k+1}\left(2\check{F}_{l-1}\check{G}_{k-l}+\frac{2u}{w}\check{F}_{l-1}\check{H}_{k-l}\right)+2u\check{G}_{k}\\
                  &=2\displaystyle\left(\sum_{l=0}^{k+1}\check{F}_{l-1}\check{G}_{k-l}-\sum_{l=0}^{k-1}\check{F}_{l}\check{G}_{k-1-l}\right)+\frac{2u}{w}\left(\sum_{l=1}^{k}\check{G}_{l-1}\check{G}_{k-l}+\sum_{l=0}^{k+1}\check{F}_{l-1}\check{H}_{k-l}\right)+2u\check{G}_{k}\\
                  &=2(\check{G}_{-1}\check{F}_{k}+\check{F}_{-1}\check{G}_{k})-4u\check{G}_{k}+2u\check{G}_{k}\\
                  &=2w\check{F}_{k}.
        \end{split}
\end{equation}
It can be similarly proved that
\begin{equation}
         \check{H}_{k-1,x}-2v\check{G}_{k}=-2w\check{H}_{k}.
\end{equation}
Basing on the above calculation, we have that
\begin{equation}
          \begin{split}
                  \check{G}_{k-1,x}&=-\displaystyle\frac{1}{u}\sum_{l=0}^{k-1}\check{G}_{l-1,x}\check{F}_{k-1-l}-\sum_{l=0}^{k+1}\left(\check{G}_{l-1}\check{G}_{k-l}+\frac{v}{u}\check{F}_{l-1}\check{F}_{k-l}\right)\\
                                   &=-\displaystyle\frac{1}{u}\sum_{l=0}^{k-1}(u\check{H}_{l}-v\check{F}_{l})\check{F}_{k-1-l}-\sum_{l=0}^{k+1}\left(\check{G}_{l-1}\check{G}_{k-l}+\frac{v}{u}\check{F}_{l-1}\check{F}_{k-l}\right)\\
                                   &=-\displaystyle\sum_{l=0}^{k+1}\check{F}_{l-1}\check{H}_{k-l}-\sum_{l=1}^{k}\check{G}_{l-1}\check{G}_{k-l}+u\check{H}_{k}+v\check{F}_{k}-2w\check{G}_{k}-\frac{2v}{u}\check{F}_{-1}\check{F}_{k}\\
                                   &=u\check{H}_{k}-v\check{F}_{k}.
          \end{split}
\end{equation}
Hence, $\check{F}_l$, $\check{H}_l$ and $\check{G}_l$ are equal to
$\hat{F}_l$, $\hat{H}_l$ and $\hat{G}_l$ respectively for all
$l\in\mathbb{N}_0\cup\{-1\}$. Thus we proved
\begin{equation}
         \frac{F(\lambda)}{R(\lambda)^{1/2}}\mathop{=}\limits_{\lambda\rightarrow\infty}\sum_{l=0}^{\infty}\hat{F}_{l-1}\lambda^{-l},\quad
         \frac{H(\lambda)}{R(\lambda)^{1/2}}\mathop{=}\limits_{\lambda\rightarrow\infty}\sum_{l=0}^{\infty}\hat{H}_{l-1}\lambda^{-l},\quad
         \frac{G(\lambda)}{R(\lambda)^{1/2}}\mathop{=}\limits_{\lambda\rightarrow\infty}\sum_{l=0}^{\infty}\hat{G}_{l-1}\lambda^{-l}.
\end{equation}
Considering
\begin{equation}
            R(\lambda)^{1/2}\mathop{=}\limits_{\lambda\rightarrow\infty}\sum_{l=0}^{\infty}{c}_{l-1}(\underline{\Lambda})\lambda^{n+1-l},
\end{equation}
a comparison of the coefficients of $\lambda^{-k}$ in the following
equation
\begin{equation}
            1=R(\lambda)^{1/2}\times R(\lambda)^{-1/2}\mathop{=}\limits_{\lambda\rightarrow\infty}\left(\sum_{l=0}^{\infty}{c}_{l-1}(\underline{\Lambda})\lambda^{n+1-l}\right)\left(\sum_{l=0}^{\infty}\hat{c}_{l-1}(\underline{\Lambda})\lambda^{-n-1-l}\right)
\end{equation}
yields
\begin{equation}
            \sum_{l=0}^kc_{k-l-1}(\underline{\Lambda})\hat{c}_{l-1}(\underline{\Lambda})=\delta_{k,0},\quad k\in\mathbb{N}_0.
\end{equation}
Therefore, we compute that
\begin{equation}
            \sum_{m=0}^{k+1}{c}_{k-m}(\underline{\Lambda})\hat{F}_{m-1}
            =\sum_{m=0}^{k+1}{c}_{k-m}(\underline{\Lambda})\sum_{l=0}^mF_{l-1}\hat{c}_{m-l-1}(\underline{\Lambda})
            =\sum_{l=0}^{k+1}F_{l-1}\sum_{p=0}^{k+1-l}c_{k-l-p}(\underline{\Lambda})\hat{c}_{p-1}(\underline{\Lambda})
            =F_k,
\end{equation}
where $k=-1,\ldots,n$.\quad\quad\quad$\Box$

\section{Dubrovin-type equations and straightening out of the
flows}\setcounter{equation}{0} \hspace{0.5cm} In this section, we
introduce elliptic variables and Abel-Jacobi coordinates. Then we
derive the system of Dubrovin-type differential equations. The
straightening out of various flows is exactly given through the
Abel-Jacobi coordinates. Noticing (3.5), we write $F$ and $H$ as
finite products which take the form
\begin{equation}
         F=u\prod\limits_{j=1}^{n+1}(\lambda-\mu_{j}),\quad
         H=v\prod\limits_{j=1}^{n+1}(\lambda-\nu_{j}),\\
\end{equation}
where $\{\mu_{j}\}_{j=1}^{n+1}$ and $\{\nu_{j}\}_{j=1}^{n+1}$ are
called elliptic variables. According to the definition of
$\mathcal{K}_{n}$, we can lift the roots $\mu_{j}$ and $\nu_{j}$ to
$\mathcal{K}_{n}$ by introducing
\begin{equation}
         \hat{\mu}_{j}(x,t_{m})=\left(\mu_{j}(x,t_{m}),-G(\mu_{j}(x,t_{m}),x,t_{m})\right),
\end{equation}
\begin{equation}
         \hat{\nu}_{j}(x,t_{m})=\left(\nu_{j}(x,t_{m}),G(\nu_{j}(x,t_{m}),x,t_{m})\right),
\end{equation}
where $j=1,\ldots,n+1,$ $(x,t_{m})\in \mathbb{R}^{2}$. \\
 Noticing (3.13), we obtain
\begin{equation}
         G|_{\lambda=\mu_{k}}=\sqrt{R(\mu_{k})},\quad
         G|_{\lambda=\nu_{k}}=\sqrt{R(\nu_{k})}.
\end{equation}
By virtue of (3.3) and (4.1), we obtain
\begin{equation}
         F_{x}|_{\lambda=\mu_{k}}=-u\mu_{k,x}\prod\limits_{
                                                            \begin{subarray}{l}
                                                                 j=1\\
                                                                 j\neq k\\
                                                            \end{subarray}
                                                          }^{n+1}(\mu_{k}-\mu_{j})=-2u\mu_{k}G|_{\lambda=\mu_{k}},
\end{equation}
\begin{equation}
         H_{x}|_{\lambda=\nu_{k}}=-v\nu_{k,x}\prod\limits_{
                                                            \begin{subarray}{l}
                                                                 j=1\\
                                                                 j\neq k\\
                                                            \end{subarray}
                                                          }^{n+1}(\nu_{k}-\nu_{j})=2v\nu_{k}G|_{\lambda=\nu_{k}}.
\end{equation}
From (4.4)-(4.6) we have
\begin{equation}
      \begin{split}
             \mu_{k,x}=\frac{2\mu_{k}\sqrt{R(\mu_{k})}}{\prod\limits_{
                                                                     \begin{subarray}{l}
                                                                            j=1\\
                                                                            j\neq k\\
                                                                     \end{subarray}
                                                                    }^{n+1}(\mu_{k}-\mu_{j})},\quad
             \nu_{k,x}=\frac{-2\nu_{k}\sqrt{R(\nu_{k})}}{\prod\limits_{
                                                                     \begin{subarray}{l}
                                                                            j=1\\
                                                                            j\neq k\\
                                                                     \end{subarray}
                                                                    }^{n+1}(\nu_{k}-\nu_{j})},\quad 1\leq k\leq n+1.
      \end{split}
\end{equation}
Similarly, we get the evolution of $\{\mu_{j}\}$ and $\{\nu_{j}\}$
along the $t_{m}$-flow
\begin{equation}
      \begin{split}
             \mu_{k,t_{m}}=\frac{2V_{12}^{(m)}(\mu_{k})\sqrt{R(\mu_{k})}}{u\prod\limits_{
                                                                                          \begin{subarray}{l}
                                                                                                j=1\\
                                                                                                j\neq k\\
                                                                                          \end{subarray}
                                                                                        }^{n+1}(\mu_{k}-\mu_{j})},\quad
             \nu_{k,t_{m}}=\frac{-2V_{21}^{(m)}(\nu_{k})\sqrt{R(\nu_{k})}}{v\prod\limits_{
                                                                                          \begin{subarray}{l}
                                                                                                j=1\\
                                                                                                 j\neq k\\
                                                                                          \end{subarray}
                                                                                         }^{n+1}(\nu_{k}-\nu_{j})},\quad 1\leq k\leq n+1.
      \end{split}
\end{equation}

In order to straighten out of the corresponding flows, we equip
$\mathcal{K}_{n}$ with canonical basis cycles:
$\tilde{a}_{1},\ldots,\tilde{a}_{n}$;
$\tilde{b}_{1},\ldots,\tilde{b}_{n}$, which are independent and have
intersection numbers as follows
\begin{equation}
        \tilde{a}_{j}\circ \tilde{a}_{k}=0,\quad
        \tilde{b}_{j}\circ\tilde{b}_{k}=0,\quad
        \tilde{a}_{j}\circ \tilde{b}_{k}=\delta_{jk}.
\end{equation}
For the present, we will choose our basis as the following
set$^{[8]}$
\begin{equation}
        \tilde{\omega}_{l}=\frac{\lambda^{l-1}d\lambda}{\sqrt{R(\lambda)}},\quad 1\leq l\leq n,
\end{equation}
which are $n$ linearly independent homomorphic differentials on
$\mathcal{K}_{n}$. Then the period matrices $A$ and $B$ can be
constructed from
\begin{equation}
        A_{kj}=\int_{\tilde{a}_{j}}\tilde{\omega}_{k},\quad
        B_{kj}=\int_{\tilde{b}_{j}}\tilde{\omega}_{k}.
\end{equation}
It is possible to show that matrices $A$ and $B$ are
invertible [3]. Now we define the matrices $C$ and $\tau$ by
$C=A^{-1}$, $\tau=A^{-1}B$. The matrix $\tau$ can be shown to be
symmetric ($\tau_{kj}=\tau_{jk}$), and it has positive definite
imaginary part (Im$\tau>0$). If we normalize $\tilde{\omega}_{l}$
into the new basis $\omega_{j}$,
\begin{equation}
        \omega_{j}=\sum\limits_{l=1}^{n}C_{jl}\tilde{\omega}_{l},\quad 1\leq j\leq n,
\end{equation}
then we obtain
\begin{equation}
        \int_{\tilde{a}_{k}}\omega_{j}=\sum\limits_{l=1}^{n}C_{jl}\int_{\tilde{a}_{k}}\tilde{\omega}_{l}=\delta_{jk},\int_{\tilde{b}_{k}}\omega_{j}=\tau_{jk}.
\end{equation}
Let $\mathcal{T}_{n}$ be the period lattice
$\mathcal{T}_{n}=\{\underline{z}\in\mathbb{C}^{n}|\underline{n}+\tau
\underline{m},\underline{m},\underline{n}\in\mathbb{Z}^{n}\}$. The
complex torus $\mathscr{T}=\mathbb{C}^{n}/\mathcal{T}_{n}$
  is called the Jacobian variety of
$\mathcal{K}_{n}$. Now we introduce the Abel map
$\mathcal{A}(P):\mathrm{Div}(\mathcal{K}_{n})\rightarrow
\mathscr{T}$
\begin{equation}
         \mathcal{A}(P)=\left(\int_{P_{0}}^{P}\underline{\omega}\right)(\mathrm{mod}\mathcal{T}_{n}),\quad
         \mathcal{A}\left(\sum n_{k}P_{k}\right)=\sum n_{k}\mathcal{A}(P_{k}),
\end{equation}
where $P$, $P_{k}\in\mathcal{K}_{n}$,
$\underline{\omega}=(\omega_{1},\ldots,\omega_{n})$. Considering two
special divisors $\sum\limits_{k=1}^{n+1}P_{k}^{(l)},~l=1,2,$ we
define the Abel-Jacobi coordinates as follows
\begin{equation}
         \mathcal{A}\left(\sum_{k=1}^{n+1}P_{k}^{(l)}\right)=\sum_{k=1}^{n+1}\mathcal{A}(P_{k}^{(l)})=\sum_{k=1}^{n+1}\int_{P_{0}}^{P_{k}^{(l)}}\underline{\omega}=\underline{\rho}^{(l)},
\end{equation}
with $P_{k}^{(1)}=\hat{\mu}_{k}(x,t_{m}),$ and
$P_{k}^{(2)}=\hat{\nu}_{k}(x,t_{m}),$ whose components are
\begin{equation}
         \sum_{k=1}^{n+1}\int_{P_{0}}^{P_{k}^{(l)}}\omega_{j}=\rho^{(l)}_{j},\quad 1\leq j\leq n,\quad l=1,2.
\end{equation}
Without loss of generality, we choose the branch point
$P_{0}=(\lambda_{j_{0}},0)$, $j_{0}\in \{1,\ldots,2n+2\}$, as a
convenient base point, and $\lambda(P_{0})$ is its local coordinate.
From (4.7), we get
\begin{equation}
         \partial_{x}\rho_{j}^{(1)}=\sum\limits_{l=1}^{n}\sum\limits_{k=1}^{n+1}C_{jl}\frac{\mu_{k}^{l-1}\mu_{k,x}}{\sqrt{R(\mu_{k})}}
                                   =\sum\limits_{l=1}^{n}\sum\limits_{k=1}^{n+1}\frac{2C_{jl}\mu_{k}^{l}}{\prod\limits_{
                                                                                                                        \begin{subarray}{l}
                                                                                                                                  r=1\\
                                                                                                                                  r\neq k\\
                                                                                                                        \end{subarray}
                                                                                                                       }^{n+1}(\mu_{k}-\mu_{r})}
                                   =2C_{jn},\quad 1\leq j\leq n,
\end{equation}
\begin{equation}
         \partial_{x}\rho_{j}^{(2)}=\sum\limits_{l=1}^{n}\sum\limits_{k=1}^{n+1}C_{jl}\frac{\nu_{k}^{l-1}\nu_{k,x}}{\sqrt{R(\nu_{k})}}
                                   =\sum\limits_{l=1}^{n}\sum\limits_{k=1}^{n+1}\frac{-2C_{jl}\nu_{k}^{l}}{\prod\limits_{
                                                                                                                        \begin{subarray}{l}
                                                                                                                                  r=1\\
                                                                                                                                  r\neq k\\
                                                                                                                        \end{subarray}
                                                                                                                       }^{n+1}(\nu_{k}-\nu_{r})}
                                   =-2C_{jn},\quad 1\leq j\leq n,
\end{equation}
with the aid of the following equalities:
\begin{equation}
      \sum\limits_{k=1}^{n}\frac{\mu_{k}^{l-1}}{\prod\limits_{r\neq k}(\mu_{k}-\mu_{r})}
                      =\left\{
                         \begin{array}{ll}
                            \delta_{ln},\quad 1\leq l\leq n,\\
                            \sum\limits_{r_{1}+\ldots+r_{n}=l-n,r_{j}\geq 0}\mu_{1}^{r_{1}}\cdots \mu_{n}^{r_{n}},\quad l> n.
                         \end{array}
                      \right.
\end{equation}

\textbf{Theorem 4.1.} (Straightening out of the $t_m-$flow)
\begin{equation}
     \partial_{t_{m}}\underline{\rho}^{(1)}=2\sum_{l=0}^{m}\beta_{l-1}\underline{C}_{n-m+l},
\end{equation}
\begin{equation}
     \partial_{t_{m}}\underline{\rho}^{(2)}=-2\sum_{l=0}^{m}\beta_{l-1}\underline{C}_{n-m+l},
\end{equation}
where
    $\underline{\rho}^{(i)}=({\rho}^{(i)}_1,\ldots,{\rho}^{(i)}_{n})$,
    $\underline{C}_k=(C_{1k},\ldots,C_{nk})$, $1\leq k\leq n$,
and the recursive formula:
\begin{equation}
      \begin{split}
          \beta_{-1}=1,\quad
          \beta_{0}=-\alpha_0,\quad
          \beta_{1}=\alpha_0^2-\alpha_1,\quad
          \beta_k=-\sum\limits_{j=0}^{k}\alpha_j\beta_{k-1-j}.
      \end{split}
\end{equation}

{\textbf Proof.}  Here we only give the proof of (4.20). Equation
(4.21) can be proved in a similar way. Using (3.10), we arrive at
\begin{equation}
      F_k=\sum_{j=0}^{k+1}\alpha_{j-1}V_{12,k-j}^{(m)},
\end{equation}
which implies
\begin{equation}
      V_{12,k}^{(m)}=\sum_{j=0}^{k+1}\beta_{j-1}F_{k-j}.
\end{equation}

In fact, it is easy to see that $F_{-1}=V_{12,-1}^{(m)}=u.$ Suppose
that (4.24) holds. Then a direct calculation shows by (4.23) that
\begin{equation}
       \begin{split}
              V_{12,k+1}^{(m)}&=F_{k+1}-\alpha_0V_{12,k}^{(m)}-\ldots-\alpha_kV_{12,0}^{(m)}-\alpha_{k+1}V_{12,-1}^{(m)}\\
                              &=F_{k+1}-\alpha_0(\beta_{-1}F_k+\beta_0F_{k-1}+\ldots+\beta_kF_{-1})-\ldots-\alpha_k(\beta_{-1}F_0+\beta_0F_{-1})-\alpha_{k+1}F_{-1}\\
                              &=F_{k+1}+(-\alpha_0\beta_{-1})F_k+(-\alpha_0\beta_0-\alpha_1\beta_{-1})F_{k-1}+\ldots+(-\alpha_0\beta_k-\ldots-\alpha_{k+1})F_{-1}\\
                              &=F_{k+1}+\beta_0F_k+\beta_1F_{k-1}+\ldots+\beta_{k+1}F_{-1}\\
                              &=\sum\limits_{j=0}^{k+2}\beta_{j-1}F_{k+1-j}.
       \end{split}
\end{equation}
Therefore (4.24) holds. From (4.8), (4.15), (4.19) and (4.24), we
have
\begin{equation}
       \begin{split}
              \partial_{t_{m}}\rho_j^{(1)}&=\displaystyle\sum_{l=1}^{n}\sum_{k=1}^{n+1}\frac{C_{jl}\mu^{l-1}_k\mu_{k,t_m}}{\sqrt{R(\mu_k)}}
                                           =\sum_{l=1}^{n}\sum_{k=1}^{n+1}\frac{2C_{jl}\mu^{l-1}_kV_{12}^{(m)}(\mu_k)}{u\prod\limits_{\stackrel{r=1}{r\neq k}}^{n+1}(\mu_k-\mu_r)}\\
                                          &=\displaystyle\sum_{l=1}^{n}\sum_{k=1}^{n+1}\frac{2C_{jl}\mu^{l-1}_k}{u\prod\limits_{\stackrel{r=1}{r\neq k}}^{n+1}(\mu_k-\mu_r)}\left(\displaystyle\sum_{p=0}^mV^{(m)}_{12,p-1}\mu_k^{m+1-p}\right)\\
                                          &=\displaystyle\sum_{l=1}^{n}\sum_{k=1}^{n+1}\frac{2C_{jl}}{u\prod\limits_{\stackrel{r=1}{r\neq k}}^{n+1}(\mu_k-\mu_r)}\sum_{p=0}^m\left(\displaystyle\sum_{q=0}^p\beta_{q-1}F_{p-1-q}\right)\mu_k^{m+l-p}\\
                                          &=\displaystyle\sum_{q=0}^m\frac{2\beta_{q-1}}{u}\sum_{p=q}^mF_{p-1-q}\sum_{l=0}^{m-p}C_{j,n-(m-p)+l}\Gamma_l\\
                                          &=\displaystyle\sum_{q=0}^m\frac{2\beta_{q-1}}{u}\sum_{k=0}^{m-q}\sum_{l=0}^kC_{j,n-(m-q)+k}F_{l-1}\Gamma_{k-l},
       \end{split}
\end{equation}
with
\begin{equation}
       \Gamma_0=1,\quad
       \Gamma_k=\displaystyle\sum_{\stackrel{j_1+\cdots+j_{n+1}=k}{j_i\geq 0}}\mu_{1}^{j_{1}}\ldots\mu_{n+1}^{j_{n+1}},\quad k\geq 1.
\end{equation}\\
Therefore, we obtain that
\begin{equation}
      \begin{split}
            \partial_{t_{m}}\rho_j^{(1)}=\displaystyle\sum_{l=0}^{m}\frac{2\beta_{l-1}}{u}F_{-1}C_{j,n-m+l}
                                        =2\displaystyle\sum_{l=0}^{m}\beta_{l-1}C_{j,n-m+l}
      \end{split}
\end{equation}
in view of the formula [21]
\begin{equation}
      \begin{split}
            \displaystyle\sum_{\stackrel{j_1+j_2=k}{j_i\geq 0}}\Gamma_{j_1}F_{j_2-1}=0,\quad 1\leq k\leq m,
      \end{split}
\end{equation}
where
\begin{equation}
       F_{-1}=u,\quad
       F_{0}=\displaystyle -u\sum_{j=1}^{n+1}\mu_{j},\quad
       F_l=\displaystyle (-1)^{l+1}u\sum_{\stackrel{j_1<\cdots<j_{l+1}}{j_i\geq 1}}\mu_{j_{1}}\ldots\mu_{j_{l+1}},\quad 0\leq l\leq n.
\end{equation}
This completes the proof of the theorem.$\quad\quad\quad\Box$

\section{Quasi-periodic solutions}\setcounter{equation}{0} \hspace{0.5cm}
In the section, we shall construct quasi-periodic solutions of the
Heisenberg hierarchy (2.11). From (3.13) and (3.14) we have
\begin{equation}
       y^{2}=G^{2}+FH,
\end{equation}
that is
\begin{equation}
      (y-G)(y+G)=FH,
\end{equation}
and then we can define the meromorphic function $\phi(P,x,t_{m})$ on
$\mathcal{K}_{n}$
\begin{equation}
      \phi(P,x,t_{m})=\frac{y-G}{F}=\frac{H}{y+G},
\end{equation}
where $P=(\lambda,y)\in \mathcal{K}_{n}\backslash
\{P_{\infty+},P_{\infty-}\}$.

\textbf{Lemma 5.1.} Suppose that
$u(x,t_{m}),v(x,t_{m}),w(x,t_{m})\in C^{\infty}(\mathbb{R}^{2})$
satisfy the hierarchy (2.11). Let $\lambda_{j}\in
\mathbb{C}\backslash\{0\}$, $1\leq j\leq 2n+2$, and
$P=(\lambda,y)\in \mathcal{K}_{n}\backslash
\{P_{\infty+},P_{\infty-}\}$. Then
\begin{equation}
       \phi\mathop{=}\limits_{\zeta\rightarrow 0}\left\{
                                                     \begin{split}
                                                            -\frac{1+w}{u}+\frac{(1+w)u_{x}-uw_{x}}{2u^{2}}\zeta+O(\zeta^{2}),~~as~~P\rightarrow P_{\infty+},\\
                                                            \frac{1-w}{u}+\frac{(1-w)u_{x}+uw_{x}}{2u^{2}}\zeta+O(\zeta^{2}),~~as~~P\rightarrow P_{\infty-},\\
                                                     \end{split}
                                                 \right.\qquad\zeta=\lambda^{-1}.
\end{equation}
\indent\textbf{Proof.}  According to Lemma 3.1, we have
\begin{equation}
             y=\mp\prod\limits_{j=1}^{2n+2}(\lambda-\lambda_{j})^{\frac{1}{2}}\mathop{=}\limits_{\zeta\rightarrow 0}\mp\zeta^{-n-1}\left(1+\alpha_{0}\zeta+O(\zeta^{2})\right),~~as~~P\rightarrow P_{\infty\pm}.
\end{equation}
From (3.5), we obtain
\begin{equation}
       \begin{split}
             F^{-1}&\mathop{=}\limits_{\zeta\rightarrow 0}\zeta^{n+1}(F_{-1}+F_{0}\zeta+O(\zeta^{2}))^{-1}\\
                   &\mathop{=}\limits_{\zeta\rightarrow 0}\zeta^{n+1}(u^{-1}-u^{-2}F_{0}\zeta+O(\zeta^{2})),~~as~~P\rightarrow P_{\infty\pm},
       \end{split}
\end{equation}
and
\begin{equation}
       G\mathop{=}\limits_{\zeta\rightarrow 0}-\frac{1}{2}w\zeta^{-n-1}(g_{-1}+g_{0}\zeta+O(\zeta^{2})),~~as~~P\rightarrow P_{\infty\pm}.
\end{equation}
Then according to the definition of $\phi$ in (5.3), we have
\begin{equation}
      \begin{split}
            \phi&=\frac{y-G}{F}\\
                &\mathop{=}\limits_{\zeta\rightarrow 0}\left(\mp \left(1+\alpha_{0}\zeta+O(\zeta^{2})\right)+\frac{1}{2}w\left(g_{-1}+g_{0}\zeta+O(\zeta^{2})\right)\right)\left(u^{-1}-u^{-2}F_{0}\zeta+O(\zeta^{2})\right)\\
                &\mathop{=}\limits_{\zeta\rightarrow 0}\left\{
                                                             \begin{split}
                                                                    -\frac{1+w}{u}+\frac{(1+w)u_{x}-uw_{x}}{2u^{2}}\zeta+O(\zeta^{2}),~~as~~P\rightarrow P_{\infty+},\\
                                                                    \frac{1-w}{u}+\frac{(1-w)u_{x}+uw_{x}}{2u^{2}}\zeta+O(\zeta^{2}),~~as~~P\rightarrow P_{\infty-},\\
                                                             \end{split}
                                                       \right.
      \end{split}
\end{equation}
which proves this lemma.\quad\quad\quad$\Box$\\
Hence the divisor of $\phi(P,x,t_{m})$ is
\begin{equation}
      (\phi(P,x,t_{m}))=D_{\hat{\nu}_{1}(x,t_{m}),\ldots,\hat{\nu}_{n+1}(x,t_{m})}-D_{\hat{\mu}_{1}(x,t_{m}),\ldots,\hat{\mu}_{n+1}(x,t_{m})}.
\end{equation}

Let
$\omega^{(3)}_{\hat{\nu}_{n+1}(x,t_{m}),\hat{\mu}_{n+1}(x,t_{m})}(P)$
denote the normalized Abelian differentials of the third kind
holomorphic on $\mathcal{K}_{n}\backslash \
\{\hat{\nu}_{n+1}(x,t_{m}),\hat{\mu}_{n+1}(x,t_{m})\}$ with simple
poles at $\hat{\nu}_{n+1}(x,t_{m})$ and $\hat{\mu}_{n+1}(x,t_{m})$
with residues $\pm1$, respectively, which can be expressed as
\begin{equation}
       \begin{split}
              \omega^{(3)}_{\hat{\nu}_{n+1}(x,t_{m}),\hat{\mu}_{n+1}(x,t_{m})}(P)=&\left(\frac{y+G(\nu_{n+1}(x,t_{m}))}{\lambda-\nu_{n+1}(x,t_{m})}-\frac{y-G(\mu_{n+1}(x,t_{m}))}{\lambda-\mu_{n+1}(x,t_{m})}\right)\frac{d\lambda}{2y}\\
                                                                                  &+\frac{\gamma_{n}}{y}\prod\limits_{j=1}^{n-1}(\lambda-\gamma_{j})d\lambda,
       \end{split}
\end{equation}
where  $\gamma_{j}\in \mathbb{C}$, $j=1,\ldots,n$, are constants
that are determined by
\begin{equation}
       \int_{\tilde{a}_{j}}\omega^{(3)}_{\hat{\nu}_{n+1}(x,t_{m}),\hat{\mu}_{n+1}(x,t_{m})}(P)=0,\quad j=1,\ldots,n.
\end{equation}
The explicit formula (5.10) then implies
\begin{equation}
      \begin{split}
            \omega^{(3)}_{\hat{\nu}_{n+1}(x,t_{m}),\hat{\mu}_{n+1}(x,t_{m})}(P)\mathop{=}\limits_{\zeta\rightarrow 0}\left\{
                                                                                                                    \begin{array}{ll}
                                                                                                                          (\zeta^{-1}+O(1))d\zeta, ~~as~~P\rightarrow \hat{\nu}_{n+1}(x,t_{m}),\zeta=\lambda-\nu_{n+1}(x,t_{m}) \\
                                                                                                                          (M(x,t_{m})\pm\gamma_{n})d\zeta, ~~as~~P\rightarrow P_{\infty\pm},\zeta=\lambda^{-1} \\
                                                                                                                          (-\zeta^{-1}+O(1))d\zeta, ~~as~~P\rightarrow
                                                                                                                          \hat{\mu}_{n+1}(x,t_{m}),\zeta=\lambda-\mu_{n+1}(x,t_{m}),
                                                                                                                    \end{array}
                                                                                                               \right.
     \end{split}
\end{equation}
where
$M(x,t_{m})=\frac{1}{2}(\mu_{n+1}(x,t_{m})-\nu_{n+1}(x,t_{m}))$.
Therefore,
\begin{equation}
     \begin{split}
             \int_{P_{0}}^{P}\omega^{(3)}_{\hat{\nu}_{n+1}(x,t_{m}),\hat{\mu}_{n+1}(x,t_{m})}(P)&\mathop{=}\limits_{\zeta\rightarrow 0}\left\{
                                                                                                                    \begin{array}{ll}
                                                                                                                           \mathrm{ln}\zeta+\omega_0(\hat{\nu}_{n+1}(x,t_{m}))+O(\zeta),~~as~~P\rightarrow \hat{\nu}_{n+1}(x,t_{m}), \\
                                                                                                                           \omega_{0}^{\infty\pm}+(M(x,t_{m})\pm\gamma_{n})\zeta+O(\zeta^{2}), ~~as~~P\rightarrow P_{\infty\pm},\\
                                                                                                                           -\mathrm{ln}\zeta+\omega_0(\hat{\mu}_{n+1}(x,t_{m}))+O(\zeta),~~as~~P\rightarrow \hat{\mu}_{n+1}(x,t_{m}), \\
                                                                                                                    \end{array}
                                                                                                                \right.
    \end{split}
\end{equation}
where $\omega_0(\hat{\nu}_{n+1}(x,t_{m}))$,
 $\omega_0(\hat{\mu}_{n+1}(x,t_{m}))$,
  $\omega_{0}^{\infty\pm}$
are integration constants.

The Riemann theta function [3] is defined as
\begin{equation}
         \theta(\underline{z}(P,D))=\theta(\underline{K}-\mathcal{A}(P)+\mathcal{A}(D)),
\end{equation}
 where
$P\in\mathcal{K}_{n}$, $D\in\mathrm{Div}(\mathcal{K}_{n})$, and
$\underline{K}=(K_{1},\ldots,K_{n})$ is defined by
\begin{equation}
         K_{j}=\frac{1}{2}(1+\tau_{jj})-\sum_{\begin{subarray}{l}
                                                      k=1\\
                                                      k\neq j\\
                                               \end{subarray}}^{n}\int_{\tilde{a}_{k}}\omega_{k}\int_{P_{0}}^{P}\omega_{j},\quad j=1,\ldots,n.
\end{equation}
Denote
\begin{equation}
        \underline{\varrho}^{(l)}=\sum_{k=1}^{n}\int_{P_{0}}^{P_{k}^{(l)}}\underline{\omega}
\end{equation}
with $P_{k}^{(1)}=\hat{\mu}_{k}(x,t_{m}),$ and
$P_{k}^{(2)}=\hat{\nu}_{k}(x,t_{m})$, $l=1,2.$ Then we have
\begin{equation}
        \theta(\underline{z}(P,D_{\underline{\hat{\mu}}(x,t_{m})}))=\theta(\underline{K}-\mathcal{A}(P)+\underline{\varrho}^{(1)}),
\end{equation}
\begin{equation}
        \theta(\underline{z}(P,D_{\underline{\hat{\nu}}(x,t_{m})}))=\theta(\underline{K}-\mathcal{A}(P)+\underline{\varrho}^{(2)}),
\end{equation}
where
     $D_{\hat{\underline{\mu}}(x,t_{m})}=\sum\limits_{j=1}^{n}\hat{\mu}_{j}(x,t_{m})$,
     $D_{\hat{\underline{\nu}}(x,t_{m})}=\sum\limits_{j=1}^{n}\hat{\nu}_{j}(x,t_{m}).$

\textbf{Theorem 5.1.} Let $P=(\lambda,y)\in
\mathcal{K}_{n}\backslash \{P_{\infty+},P_{\infty-}\}$,
$(x,t_{m})\in \Omega$, where $\Omega\subseteq \mathbb{R}^{2}$ is
open and connected. Suppose $u(x,t_{m}),v(x,t_{m}),w(x,t_{m})\in
C^{\infty}(\Omega)$ satisfy the hierarchy of (2.11), and assume that
$\lambda_{j}$, $1\leq j\leq 2n+2$, in (3.13) satisfy $\lambda_{j}\in
\mathbb{C}\backslash\{0\}$, and $\lambda_{j}\neq\lambda_{k}$ as
$j\neq k$. Moreover, suppose that
$D_{\hat{\mu}_{1}(x,t_{m}),\ldots,\hat{\mu}_{n+1}(x,t_{m})}$, or
equivalently,
$D_{\hat{\nu}_{1}(x,t_{m}),\ldots,\hat{\nu}_{n+1}(x,t_{m})}$
 is nonspecial for
$(x,t_{m})\in \Omega$. Then $u,w$ admit the following representation
\begin{equation}
      \begin{split}
            w=&\quad\left(\mathrm{exp}(\omega_{0}^{\infty+})\theta(\underline{z}(P_{\infty+},D_{\underline{\hat{\nu}}(x,t_{m})}))\theta(\underline{z}(P_{\infty-},D_{\underline{\hat{\mu}}(x,t_{m})}))\right.\\
              &\quad\left.+\exp(\omega_{0}^{\infty-})\theta(\underline{z}(P_{\infty+},D_{\underline{\hat{\mu}}(x,t_{m})}))\theta(\underline{z}(P_{\infty-},D_{\underline{\hat{\nu}}(x,t_{m})}))\right)\\
              &\div\left(\mathrm{exp}(\omega_{0}^{\infty+})\theta(\underline{z}(P_{\infty+},D_{\underline{\hat{\nu}}(x,t_{m})}))\theta(\underline{z}(P_{\infty-},D_{\underline{\hat{\mu}}(x,t_{m})}))\right.\\
              &\quad\left.-\exp(\omega_{0}^{\infty-})\theta(\underline{z}(P_{\infty+},D_{\underline{\hat{\mu}}(x,t_{m})}))\theta(\underline{z}(P_{\infty-},D_{\underline{\hat{\nu}}(x,t_{m})}))\right),
      \end{split}
\end{equation}
\begin{equation}
        \begin{split}
              \frac{u_{x}}{u}+\frac{ww_{x}}{1-w^{2}}=&\quad\frac{\sum_{j=1}^{n}C_{jn}\partial_{\sigma_{j}}\theta(\underline{K}-\mathcal{A}(P_{\infty-})+\underline{\varrho}^{(2)}+\underline{\sigma})|_{\underline{\sigma}=0}}{\theta(\underline{z}(P_{\infty-},D_{\underline{\hat{\nu}}(x,t_{m})}))}\\
                                                     &+\frac{\sum_{j=1}^{n}C_{jn}\partial_{\sigma_{j}}\theta(\underline{K}-\mathcal{A}(P_{\infty+})+\underline{\varrho}^{(2)}+\underline{\sigma})|_{\underline{\sigma}=0}}{\theta(\underline{z}(P_{\infty+},D_{\underline{\hat{\nu}}(x,t_{m})}))}\\
                                                     &-\frac{\sum_{j=1}^{n}C_{jn}\partial_{\sigma_{j}}\theta(\underline{K}-\mathcal{A}(P_{\infty-})+\underline{\varrho}^{(1)}+\underline{\sigma})|_{\underline{\sigma}=0}}{\theta(\underline{z}(P_{\infty-},D_{\underline{\hat{\mu}}(x,t_{m})}))}\\
                                                     &-\frac{\sum_{j=1}^{n}C_{jn}\partial_{\sigma_{j}}\theta(\underline{K}-\mathcal{A}(P_{\infty+})+\underline{\varrho}^{(1)}+\underline{\sigma})|_{\underline{\sigma}=0}}{\theta(\underline{z}(P_{\infty+},D_{\underline{\hat{\mu}}(x,t_{m})}))}-2\gamma_{n}.
        \end{split}
\end{equation}

\textbf{Proof.}  We introduce the local coordinate
$\zeta=\lambda^{-1}$ near $P_{\infty\pm}$. From the definition
(4.12) of the normalized bases $\omega_j$, we have that
\begin{equation}
      \begin{split}
            \underline{\omega}&=(\omega_{1},\omega_{2},\ldots,\omega_{n})
                               =\mp\sum\limits_{l=1}^{n}\underline{C}_{l}\frac{\lambda^{l-1}d\lambda}{\prod_{j=1}^{2n+2}(\lambda-\lambda_{j})^{\frac{1}{2}}}\\
                              &=\pm\sum\limits_{l=1}^{n}\underline{C}_{l}\zeta^{n-l}\prod_{j=1}^{2n+2}(1-\lambda_{j}\zeta)^{-\frac{1}{2}}d\zeta
                               \mathop{=}\limits_{\zeta\rightarrow 0}\pm(\underline{C}_{n}+O(\zeta))d\zeta, ~as~P\rightarrow P_{\infty\pm}.
      \end{split}
\end{equation}
According to Riemann's vanishing theorem [15], the definition and
asymptotic properties of $\phi$, $\phi$ has expression of the
following type
\begin{equation}
     \phi(P,x,t_{m})=N(x,t_{m})\frac{\theta(\underline{z}(P,D_{\underline{\hat{\nu}}(x,t_{m})}))}{\theta(\underline{z}(P,D_{\underline{\hat{\mu}}(x,t_{m})}))}
                     \mathrm{exp}\left(\int_{P_{0}}^{P}\omega^{(3)}_{\hat{\nu}_{n+1}(x,t_{m}),\hat{\mu}_{n+1}(x,t_{m})}(P)\right),
\end{equation}
where $N(x,t_{m})$ is independent of $P\in\mathcal{K}_{n}$. Given
(5.13), we can derive that
\begin{equation}
     \begin{split}
            \mathrm{exp}\left(\int_{P_{0}}^{P}\omega^{(3)}_{\hat{\nu}_{n+1}(x,t_{m}),\hat{\mu}_{n+1}(x,t_{m})}(P)\right)
            \mathop{=}\limits_{\zeta\rightarrow 0}\mathrm{exp}(w_{0}^{\infty\pm})(1+(M(x,t_{m})\pm\gamma_{n})\zeta+O(\zeta^{2})),~as~P\rightarrow P_{\infty\pm}.
     \end{split}
\end{equation}
Combining (5.17), (5.18) and (5.21), we obtain the following
asymptotic expansion
\begin{equation}
      \begin{split}
             &\frac{\theta(\underline{z}(P,D_{\underline{\hat{\nu}}(x,t_{m})}))}{\theta(\underline{z}(P,D_{\underline{\hat{\mu}}(x,t_{m})}))}
              =\frac{\theta(\underline{K}-\mathcal{A}(P_{\infty+})+\underline{\varrho}^{(2)}+\mathcal{A}(P_{\infty+})-\mathcal{A}(P))}{\theta(\underline{K}-\mathcal{A}(P_{\infty+})+\underline{\varrho}^{(1)}+\mathcal{A}(P_{\infty+})-\mathcal{A}(P))}\\
             &\mathop{=}\limits_{\zeta\rightarrow 0}\frac{\theta(\underline{K}-\mathcal{A}(P_{\infty+})+\underline{\varrho}^{(2)}-\underline{C}_{n}\zeta+O(\zeta^{2}))}{\theta(\underline{K}-\mathcal{A}(P_{\infty+})+\underline{\varrho}^{(1)}-\underline{C}_{n}\zeta+O(\zeta^{2}))}\\
             &\mathop{=}\limits_{\zeta\rightarrow 0}\frac{\theta(\underline{z}(P_{\infty+},D_{\underline{\hat{\nu}}(x,t_{m})}))-\sum_{j=1}^{n}C_{jn}\partial_{\sigma_{j}}\theta(\underline{K}-\mathcal{A}(P_{\infty+})+\underline{\varrho}^{(2)}+\underline{\sigma})|_{\underline{\sigma}=0}\zeta+O(\zeta^{2})}
                        {\theta(\underline{z}(P_{\infty+},D_{\underline{\hat{\mu}}(x,t_{m})}))-\sum_{j=1}^{n}C_{jn}\partial_{\sigma_{j}}\theta(\underline{K}-\mathcal{A}(P_{\infty+})+\underline{\varrho}^{(1)}+\underline{\sigma})|_{\underline{\sigma}=0}\zeta+O(\zeta^{2})}\\
             &\mathop{=}\limits_{\zeta\rightarrow 0}\frac{\theta(\underline{z}(P_{\infty+},D_{\underline{\hat{\nu}}(x,t_{m})}))}{\theta(\underline{z}(P_{\infty+},D_{\underline{\hat{\mu}}(x,t_{m})}))}
                        \left(1-\frac{\sum_{j=1}^{n}C_{jn}\partial_{\sigma_{j}}\theta(\underline{K}-\mathcal{A}(P_{\infty+})+\underline{\varrho}^{(2)}+\underline{\sigma})|_{\underline{\sigma}=0}}{\theta(\underline{z}(P_{\infty+},D_{\underline{\hat{\nu}}(x,t_{m})}))}\zeta+O(\zeta^{2})\right)\\
             &~~~~~\times\left(1+\frac{\sum_{j=1}^{n}C_{jn}\partial_{\sigma_{j}}\theta(\underline{K}-\mathcal{A}(P_{\infty+})+\underline{\varrho}^{(1)}+\underline{\sigma})|_{\underline{\sigma}=0}}{\theta(\underline{z}(P_{\infty+},D_{\underline{\hat{\mu}}(x,t_{m})}))}\zeta+O(\zeta^{2})\right)\\
             &\mathop{=}\limits_{\zeta\rightarrow 0}\frac{\theta(\underline{z}(P_{\infty+},D_{\underline{\hat{\nu}}(x,t_{m})}))}{\theta(\underline{z}(P_{\infty+},D_{\underline{\hat{\mu}}(x,t_{m})}))}
                        \left[1+\left(\frac{\sum_{j=1}^{n}C_{jn}\partial_{\sigma_{j}}\theta(\underline{K}-\mathcal{A}(P_{\infty+})+\underline{\varrho}^{(1)}+\underline{\sigma})|_{\underline{\sigma}=0}}{\theta(\underline{z}(P_{\infty+},D_{\underline{\hat{\mu}}(x,t_{m})}))}\right.\right.\\
             &~~~~~\left.\left.-\frac{\sum_{j=1}^{n}C_{jn}\partial_{\sigma_{j}}\theta(\underline{K}-\mathcal{A}(P_{\infty+})+\underline{\varrho}^{(2)}+\underline{\sigma})|_{\underline{\sigma}=0}}{\theta(\underline{z}(P_{\infty+},D_{\underline{\hat{\nu}}(x,t_{m})}))}\right)\zeta+O(\zeta^{2})\right],~as~P\rightarrow
             P_{\infty+}.
      \end{split}
\end{equation}
Substituting (5.23) and (5.24) into (5.22), we have
\begin{equation}
       \begin{split}
               \phi&(P,x,t_{m})\mathop{=}\limits_{\zeta\rightarrow 0}N(x,t_{m})\mathrm{exp}(w_{0}^{\infty+})\frac{\theta(\underline{z}(P_{\infty+},D_{\underline{\hat{\nu}}(x,t_{m})}))}{\theta(\underline{z}(P_{\infty+},D_{\underline{\hat{\mu}}(x,t_{m})}))}\\
                   &\times\left[1+\left(M(x,t_{m})+\gamma_{n}+\frac{\sum_{j=1}^{n}C_{jn}\partial_{\sigma_{j}}\theta(\underline{K}-\mathcal{A}(P_{\infty+})+\underline{\varrho}^{(1)}+\underline{\sigma})|_{\underline{\sigma}=0}}{\theta(\underline{z}(P_{\infty+},D_{\underline{\hat{\mu}}(x,t_{m})}))}\right.\right.\\
                   &\quad\left.\left.-\frac{\sum_{j=1}^{n}C_{jn}\partial_{\sigma_{j}}\theta(\underline{K}-\mathcal{A}(P_{\infty+})+\underline{\varrho}^{(2)}+\underline{\sigma})|_{\underline{\sigma}=0}}{\theta(\underline{z}(P_{\infty+},D_{\underline{\hat{\nu}}(x,t_{m})}))}\right)\zeta+O(\zeta^{2})\right],~as~P\rightarrow P_{\infty+},
       \end{split}
\end{equation}
Similarly, one can get
\begin{equation}
       \begin{split}
               \phi&(P,x,t_{m})\mathop{=}\limits_{\zeta\rightarrow 0}N(x,t_{m})\mathrm{exp}(w_{0}^{\infty-})\frac{\theta(\underline{z}(P_{\infty-},D_{\underline{\hat{\nu}}(x,t_{m})}))}{\theta(\underline{z}(P_{\infty-},D_{\underline{\hat{\mu}}(x,t_{m})}))}\\
                   &\times\left[1+\left(M(x,t_{m})-\gamma_{n}+\frac{\sum_{j=1}^{n}C_{jn}\partial_{\sigma_{j}}\theta(\underline{K}-\mathcal{A}(P_{\infty-})+\underline{\varrho}^{(2)}+\underline{\sigma})|_{\underline{\sigma}=0}}{\theta(\underline{z}(P_{\infty-},D_{\underline{\hat{\nu}}(x,t_{m})}))}\right.\right.\\
                   &\quad\left.\left.-\frac{\sum_{j=1}^{n}C_{jn}\partial_{\sigma_{j}}\theta(\underline{K}-\mathcal{A}(P_{\infty-})+\underline{\varrho}^{(1)}+\underline{\sigma})|_{\underline{\sigma}=0}}{\theta(\underline{z}(P_{\infty-},D_{\underline{\hat{\mu}}(x,t_{m})}))}\right)\zeta+O(\zeta^{2})\right],~as~P\rightarrow P_{\infty-},
       \end{split}
\end{equation}
from (5.21)-(5.23). Comparing (5.25) and (5.26) with (5.4), we find
\begin{equation}
       -\frac{1+w}{u}=N(x,t_{m})\mathrm{exp}(w_{0}^{\infty+})\frac{\theta(\underline{z}(P_{\infty+},D_{\underline{\hat{\nu}}(x,t_{m})}))}{\theta(\underline{z}(P_{\infty+},D_{\underline{\hat{\mu}}(x,t_{m})}))},
\end{equation}
\begin{equation}
       \frac{1-w}{u}=N(x,t_{m})\mathrm{exp}(w_{0}^{\infty-})\frac{\theta(\underline{z}(P_{\infty-},D_{\underline{\hat{\nu}}(x,t_{m})}))}{\theta(\underline{z}(P_{\infty-},D_{\underline{\hat{\mu}}(x,t_{m})}))},
\end{equation}
\begin{equation}
      \begin{split}
              \frac{(1+w)u_{x}-uw_{x}}{2u^{2}}=&N(x,t_{m})\mathrm{exp}(w_{0}^{\infty+})\frac{\theta(\underline{z}(P_{\infty+},D_{\underline{\hat{\nu}}(x,t_{m})}))}{\theta(\underline{z}(P_{\infty+},D_{\underline{\hat{\mu}}(x,t_{m})}))}\\
                                               &\times\left(M(x,t_{m})+\frac{\sum_{j=1}^{n}C_{jn}\partial_{\sigma_{j}}\theta(\underline{K}-\mathcal{A}(P_{\infty+})+\underline{\varrho}^{(1)}+\underline{\sigma})|_{\underline{\sigma}=0}}{\theta(\underline{z}(P_{\infty+},D_{\underline{\hat{\mu}}(x,t_{m})}))}\right.\\
                                               &\quad\quad\left.+\gamma_{n}-\frac{\sum_{j=1}^{n}C_{jn}\partial_{\sigma_{j}}\theta(\underline{K}-\mathcal{A}(P_{\infty+})+\underline{\varrho}^{(2)}+\underline{\sigma})|_{\underline{\sigma}=0}}{\theta(\underline{z}(P_{\infty+},D_{\underline{\hat{\nu}}(x,t_{m})}))}\right),
      \end{split}
\end{equation}
\begin{equation}
      \begin{split}
              \frac{(1-w)u_{x}+uw_{x}}{2u^{2}}=&N(x,t_{m})\mathrm{exp}(w_{0}^{\infty-})\frac{\theta(\underline{z}(P_{\infty-},D_{\underline{\hat{\nu}}(x,t_{m})}))}{\theta(\underline{z}(P_{\infty-},D_{\underline{\hat{\mu}}(x,t_{m})}))}\\
                                               &\times\left(M(x,t_{m})+\frac{\sum_{j=1}^{n}C_{jn}\partial_{\sigma_{j}}\theta(\underline{K}-\mathcal{A}(P_{\infty-})+\underline{\varrho}^{(2)}+\underline{\sigma})|_{\underline{\sigma}=0}}{\theta(\underline{z}(P_{\infty-},D_{\underline{\hat{\nu}}(x,t_{m})}))}\right.\\
                                               &\quad\quad\left.-\gamma_{n}-\frac{\sum_{j=1}^{n}C_{jn}\partial_{\sigma_{j}}\theta(\underline{K}-\mathcal{A}(P_{\infty-})+\underline{\varrho}^{(1)}+\underline{\sigma})|_{\underline{\sigma}=0}}{\theta(\underline{z}(P_{\infty-},D_{\underline{\hat{\mu}}(x,t_{m})}))}\right).
      \end{split}
\end{equation}
Eliminating from (5.27) and (5.28) the terms $u$ and $N(x,t_{m})$,
we have (5.19). Substituting the RHS of (5.27) into (5.29), we find
\begin{equation}
      \begin{split}
              -\frac{u_{x}}{2u}+\frac{w_{x}}{2(1+w)}=&M(x,t_{m})+\frac{\sum_{j=1}^{n}C_{jn}\partial_{\sigma_{j}}\theta(\underline{K}-\mathcal{A}(P_{\infty+})+\underline{\varrho}^{(1)}+\underline{\sigma})|_{\underline{\sigma}=0}}{\theta(\underline{z}(P_{\infty+},D_{\underline{\hat{\mu}}(x,t_{m})}))}\\
                                                     &+\gamma_{n}-\frac{\sum_{j=1}^{n}C_{jn}\partial_{\sigma_{j}}\theta(\underline{K}-\mathcal{A}(P_{\infty+})+\underline{\varrho}^{(2)}+\underline{\sigma})|_{\underline{\sigma}=0}}{\theta(\underline{z}(P_{\infty+},D_{\underline{\hat{\nu}}(x,t_{m})}))}.
      \end{split}
\end{equation}
Similarly, we have
\begin{equation}
      \begin{split}
              \frac{u_{x}}{2u}+\frac{w_{x}}{2(1-w)}=&M(x,t_{m})+\frac{\sum_{j=1}^{n}C_{jn}\partial_{\sigma_{j}}\theta(\underline{K}-\mathcal{A}(P_{\infty-})+\underline{\varrho}^{(2)}+\underline{\sigma})|_{\underline{\sigma}=0}}{\theta(\underline{z}(P_{\infty-},D_{\underline{\hat{\nu}}(x,t_{m})}))}\\
                                                    &-\gamma_{n}-\frac{\sum_{j=1}^{n}C_{jn}\partial_{\sigma_{j}}\theta(\underline{K}-\mathcal{A}(P_{\infty-})+\underline{\varrho}^{(1)}+\underline{\sigma})|_{\underline{\sigma}=0}}{\theta(\underline{z}(P_{\infty-},D_{\underline{\hat{\mu}}(x,t_{m})}))},
      \end{split}
\end{equation}
from (5.28) and (5.30). Then eliminating from (5.31) and (5.32) the
term $M(x,t_{m})$ yields (5.20).\quad\quad\quad$\Box$

\vspace{0.3cm}
\noindent{\bf Acknowledgment}

This work was supported by the National Natural Science Foundation
of China (project nos. 11331008, 11171312, and 11271337).


\begin{thebibliography}{90}

\bibitem{1} Krichever, I.M., Algebraic-geometric construction of the Zaharov-Sabat equations and their periodic solutions. Dokl. Akad. Nauk SSSR 227 (1976), 394-397.
\bibitem{2} Krichever, I.M., Integration of nonlinear equations by the methods of algebraic geometry. Funct. Anal. Appl. 11 (1977), 12-26.
\bibitem{3} Dubrovin, B.A., Theta functions and nonlinear equations. Russian Math. Surveys 36 (1981), 11-92.
\bibitem{4} Date, E., Tanaka, S., Periodic multi-soliton solutions of Korteweg-de Vries equation and Toda lattice. Progr. Theoret. Phys. Suppl. 59 (1976), 107-125.
\bibitem{5} Ma, Y.C., Ablowitz, M.J., The periodic cubic Schr\"odinger equation. Stud. Appl. Math. 65 (1981), 113-158.
\bibitem{6} Smirnov, A.O., Real finite-gap regular solutions of the Kaup-Boussinesq equation. Theoret. Math. Phys. 66 (1986), 19-31.
\bibitem{7} Previato, E., Hyperelliptic quasi-periodic and soliton solutions of the nonlinear Schr\"odinger equation. Duke Math. J. 52 (1985), 329-377.
\bibitem{8} Miller, P.D., Ercolani, N.M., Krichever, I.M., Levermore, C.D., Finite genus solutions to the Ablowitz-Ladik equations. Comm. Pure Appl. Math. 48 (1995), 1369-1440.
\bibitem{9} Krichever, I.M., Novikov, S.P., Periodic and almost-periodic potential in inverse problems. Inverse Problems 15 (1999), R117-R144.
\bibitem{10} Alber, M.S., Fedorov, Y.N., Algebraic geometrical solutions for certain evolution equations and Hamiltonian flows on nonlinear subvarieties of generalized Jacobians. Inverse Problems 17 (2001), 1017-1042.
\bibitem{11} Zhou, R.G., The finite-band solution of Jaulent-Miodek equation. J. Math. Phys. 38 (1997), 2335-2546.
\bibitem{12} Gesztesy, F., Ratneseelan, R., An alternative approach to algebro-geometric solutions of the AKNS hierarchy. Rev. Math. Phys. 10 (1998), 345-391.
\bibitem{13} Cao, C.W., Wu, Y.T., Geng, X.G., Relation between the Kadomtsev-Petviashvili equation and the Confocal involutive system. J. Math. Phys. 40 (1999), 3948-3970.
\bibitem{14} Geng, X.G., Wu, Y.T., Finite-band solutions of the classical Boussinesq-Burgers equations. J. Math. Phys. 40 (1999), 2971-2982.
\bibitem{15} Gesztesy, F., Holden, H., Soliton equations and their algebro-geometric solutions. Cambridge University Press, Cambridge, 2003.
\bibitem{16} Gesztesy, F., Holden, H., Algebro-geometric solutions of the Camassa-Holm hierarchy. Rev. Mat. Iberoam. 19 (2003), 73-142.
\bibitem{17} Geng, X.G., Cao, C.W., Decomposition of the (2+1)-dimensional Gardner equation and its quasi-periodic solutions. Nonlinearity 14 (2001), 1433-1452.
\bibitem{18} Geng, X.G., Dai, H.H., Zhu, J.Y., Decomposition of the discrete Ablowitz-Ladik hierarchy. Stud. Appl. Math. 118 (2007), 281-312.
\bibitem{19} Geng, X.G., Xue, B., Quasi-periodic solutions of mixed AKNS equations. Nonlinear Anal. 73 (2010), 3662-3674.
\bibitem{20} Geng, X.G., Wu, L.H., He, G.L., Algebro-geometric constructions of the modified Boussinesq flows and quasi-periodic solutions. Physica D 240 (2011), 1262-1288.
\bibitem{21} Zhai, Y.Y., Geng, X.G., Straightening out of the flows for the Hu hierarchy and its algebro-geometric solutions. J. Math. Anal. Appl. 397 (2013), 561-576.
\bibitem{22} Landau, L.D., Lifshitz, E.M., On the theory of the dispersion of magnetic permeability in ferromagnetic bodies. Phys. Z. Sowjetunion 8 (1935), 153-169.
\bibitem{23} Bishop, A.R., Schneider, T., Solitons in condensed matter. Springer, Berlin, 1978.
\bibitem{24} Wigen, P.E., Nonlinear phenomena and chaos in magnetic materials. World Scientific, Singapore, 1994.
\bibitem{25} Dantas, C.C., An approach to loop quantum cosmology through integrable discrete Heisenberg spin chains. Found. Phys. 43 (2013), 236-242.
\bibitem{26} Takhtajan, L.A., Integration of the continuous Heisenberg spin chain through the inverse scattering method. Phys. Lett. A 64 (1977), 235-237.
\bibitem{27} Tjon, J., Wright, J., Solitons in the continuous Heisenberg spin chain. Phy. Rev. B 15 (1977), 3470-3476.
\bibitem{28} Jevicki, A., Papanicolaou, N., Semi-classical spectrum of the continuous Heisenberg spin chain. Ann. Phys. 120 (1979), 107-128.
\bibitem{29} Quispel, G.R.W., Capel, H.W., The Anisotropic Heisenberg spin chain and the nonlinear Schr\"{o}dinger equation. Physica A 117 (1983), 76-102.
\bibitem{30} Li, Y.S., Chen, D.Y., Equivalence of three kinds of nonlinear evolution equations. Acta Math. Sin. 29 (1986), 264-271.
\bibitem{31} Choudhury, A.G., Chowdhury, A.R., Nonlocal conservation laws and supersymmetric Heisenberg spin chain. Int. J. Theor. Phys. 33 (1994), 2031-2036.
\bibitem{32} Cao, C.W., Parametric representation of the finite-band solution of the Heisenberg equation. Phys. Lett. A 184 (1994), 333-338.
\bibitem{33} Bhattacharya, N., Chowdhury, A.R., Bethe ansatz for an open Heisenberg spin chain with impurity. Int. J. Theor. Phys. 33 (1994), 679-685.
\bibitem{34} Qiao, Z.J., A finite-dimensional integrable system and the involutive solutions of the higher-order Heisenberg spin chain equations. Phys. Lett. A 186 (1994), 97-102.
\bibitem{35} Du, D.L., Complex form, reduction and Lie-Poisson structure for the nonlinearized spectral problem of the Heisenberg hierarchy. Physica A 303 (2002), 439-456.
\bibitem{36} Wang, X.G., Zanardi, P., Quantum entanglement and Bell inequalities in Heisenberg spin chains. Phys. Lett. A 301 (2002), 1-6.
\bibitem{37} Wang, J., Darboux transformation and soliton solutions for the Heisenberg hierarchy. J. Phys. A 38 (2005), 5217-5226.
\bibitem{38} Guo, B.L., Zeng, M., Su, F.Q., Periodic weak solutions for a classical one-dimensional isotropic biquadratic Heisenberg spin chain. J. Math. Anal. Appl. 330 (2007), 729-739.
\bibitem{39} Its, A.R., Korepin, V.E., Generalized entropy of the Heisenberg spin chain. Theor. and Math. Phys. 164 (2010), 1136-1139.
\bibitem{40} Li, H.Z., Tian, B., Guo, R., Xue, Y.S., Qi, F.H., Gauge transformation between the first-order nonisospectral and isospectral Heisenberg hierarchies. Appl. Math. and Compu. 218 (2012), 7694-7699.
\bibitem{41} Miszczak, J.A., Gawron, P., Pucha{\l}a, Z., Qubit flip game on a Heisenberg spin chain. Quantum Inf. Process. 11 (2012), 1571-1583.
\bibitem{42} Tu, G.Z., The trace identity, a powerful tool for constructing the Hamiltonian structure of integrable systems. J. Math. Phys. 30 (1989), 330-338.
\end{thebibliography}
\end{document}